\documentclass[useAMS,usenatbib]{mn2e}

\usepackage{times}
\usepackage{psfig}
\usepackage{graphicx}

\title[ISM linked cosmic evolution of radio galaxies]{The changing ISM of massive elliptical galaxies and
cosmic evolution of radio galaxies and quasars} 

\author[A.\ Mangalam, Gopal-Krishna, P.J.\ Wiita]{A. Mangalam$^{1}$\thanks
{E-mail:mangalam@iiap.res.in (AM); krishna@ncra.tifr.res.in (G-K);
wiita@chara.gsu.edu (PJW)}, Gopal-Krishna$^{2}$ and Paul J.
Wiita$^{3,4}$\\
$^{1}$Indian Institute of Astrophysics, Sarjapur Road, Koramangala,
Bangalore, 560034, India\\
$^{2}$National Centre for Radio Astrophysics/TIFR, Post Bag 3,
Pune University Campus, Pune, 411007, India\\
$^{3}$School of Natural Sciences, Institute for Advanced Study, 1 Einstein 
Drive, Princeton, NJ 08540, USA\\
$^{4}$Department of Physics \& Astronomy, Georgia State University, 
P.O.\ Box 4106, Atlanta, GA 30302-4106, USA}
\begin{document}

\date{Accepted . Received ; in original form 2009 April 3}

\pagerange{\pageref{firstpage}--\pageref{lastpage}} \pubyear{2009}

\maketitle
\label{firstpage}
\begin{abstract} 

The recently discovered apparent dramatic expansion in the effective radii
of massive elliptical galaxies from $z \simeq 2$ to $z \simeq 0.1$
has been interpreted in terms of either galaxy mergers or the
rapid loss of cold gas due to AGN feedback.  
In examining the latter case we have quantified the extent of the expansion, 
which is uncertain observationally, in terms of
the star formation parameters and time of the expulsion of the
cold gas.
In either case, the
large global decrease in stellar density should translate into a major
drop in the ISM density and pressure with cosmic epoch.  These cosmological changes 
are expected to
have a major influence on the gas accretion mode, which will shift
from `cold' thin disk accretion at high redshifts toward `hot' 
Bondi fed ADAF accretion at low redshifts. The  decline of angular momentum
inflow would then lead to a spin down of the black hole, for which we have 
calculated more precise time scales; a value of about 0.2 Gyr is typical 
for a $10^9 M_\odot$ central black 
hole.  These results have implications for the different cosmological 
evolutionary patterns found for the
luminosity functions of powerful and weak radio galaxies.
\end{abstract}


\section{Introduction}

Powerful radio galaxies (RGs) are known to be hosted by the most massive 
($10^{11 - 12} M_{\odot}$) elliptical galaxies (Matthews, Morgan \& Schmidt 1964) 
at all cosmic epochs (e.g., De Brueck et al.\ 2002; 
Rocca-Volmerange et al.\ 2004; Seymour et al.\ 2007; Nesvadba et al.  2007).  
Moreover, the host galaxies are 
usually found to be either isolated or located in groups as opposed to
rich clusters 
(e.g., Longair \& Seldner 
1979; Best 2004; Hardcastle, Evans \& Croston 2007).  Extensive radio and 
optical studies have confirmed the original inference 
reached from radio source counts (Longair 1966) that, as compared to the 
present epoch, the space density of powerful RGs and radio-loud quasars 
was a factor of $\sim 10^3$ higher during the `quasar era' ($z$ = 2 -- 3)  
(e.g., Dunlop \& Peacock 1990; Willott et al. 2001; Grimes et 
al.\ 2004). A similarly strong cosmic evolution is exhibited by the 
optically luminous radio-quiet quasars (RQQs) (e.g., Hartwick \& Schade 
1990; Wall et al. 2005) which too reside 
almost exclusively in massive ellipticals (e.g., Falomo et al. 2008).

In contrast, low power radio galaxies, i.e., those with  $P_{\rm 178 MHz} \le 1 \times 10^{25}$ W Hz$^{-1}$ sr$^{-1}$ and typically of Fanaroff-Riley (1974) morphology class I,exhibit much weaker cosmic evolution, amounting to only a factor of $\sim$ 10 increase in  their  abundance between $z \sim 0$ and $z \sim 3$ (e.g., Jackson \& Wall 1999; Willott et al.\ 2001). 
It has been proposed that the remarkable cosmological evolution may
be a  manifestation of a  fundamental change in the dominant gas 
 accretion mode powering the AGN 
since the quasar era, which has shifted from a predominantly `cold' thin 
disc accretion at high redshifts toward `hot' Bondi type accretion at low 
redshifts (e.g., Hardcastle et al.\ 2007; cf.\ Cao \& Rawlings 
2004). The hot accretion (dominant at low 
$z$) should funnel rather little angular momentum into the central supermassive 
black hole (SMBH), leading to the possibility of its spinning down. For 
such RGs, the ADAF accretion mode (e.g., Narayan \& Yi 1995) 
is usually invoked to explain their observed faint disc emission and 
low-excitation optical spectra. Note that while such RGs mostly have an
FR I radio morphology, this is likely to be dictated mainly by environmental 
factors, especially the ambient density and its gradient (e.g., Gopal-Krishna 
\& Wiita 1988, 2000; Snellen \& Best 2001; Hardcastle et al.\ 2007; Perucho 
\& Mart{\'i} 2007; Melliani et al. 2008). 

An observational clue linked to the hypothesis of change in the 
AGN accretion mode with cosmological epoch comes from the drastic 
decrease in the cool gas content 
of massive ellipticals since the quasar era.  
Observations of CO in nearby 
massive field ellipticals (within 25 Mpc) have yielded very low detection 
rates. Curiously, the detection rates are higher for their less luminous 
counterparts (Sage et al.\ 2007; Combes et al. 2007; also, Somerville et al.\ 2008). 
This counter-intuitive result is, however, fully consistent with the idea 
that more massive ellipticals preferentially lost their cold gas due to a
stronger AGN feedback (e.g., Sazonov et al.\ 2005; Springel et al.\ 2005; Croton et al.\ 2006; Bower et al.\ 2006; Hopkins et al.\ 2008). In contrast, 
large amounts of molecular gas in high-$z$ massive ellipticals have been 
found from CO detections (e.g., Papadopoulos et al.\ 2000; De Brueck et 
al.\ 2005; Klamer et al.\ 2005) as well as PAH detections (Lutz et al.\ 2008).

It is interesting that the fraction of 
sub-millimetre bright RGs shows a sharp decline from $> 50\%$ at $z > 2.5$ 
to $< 15\%$ at $z < 2.5$ (Reuland et al.\ 2004; 
also, Greve et al.\ 2006; Seymour et al.\  2007). This shows that in RGs the 
bulk of the expulsion of 
the cold gas and/or its conversion into stars was largely accomplished 
during the quasar era. 

The available data favour the premise that jet-driven gaseous 
outflows were common in RGs during the quasar era and that such outflows 
may have played an important role in the evolution of those massive 
galaxies (e.g., Nesvadba et al.\ 2008).  The mechanical feedback of 
$\approx$ 5\% of the QSO's bolometric power during its peak phase can 
supply the energy needed to induce such intense outflows (e.g., Granato et al.\ 2004; Hopkins et al.\ 2005).
Powerful outflows, if produced by nearly-Eddington or even
super-Eddington accretion onto black holes, have been argued to be
fundamentally responsible for: the coordinated growth of the masses of the
galactic bulge and the SMBH (e.g., Silk \& Rees 1998; Fabian 1999;  
King 2003); controlling the rate of growth of galaxies
(e.g., Bower et al.\ 2005; Croton et al.\ 2006; Best et al.\ 2006); as well as 
offsetting the cluster cooling flows via heating 
of their ICM (e.g., Binney \& Tabor 1995; Fabian et al.\ 2003; B{\^i}rzan et al.\ 2004).

Whereas the expulsion of huge quantities of cold gas from massive 
elliptical galaxies over the past $\sim$10 gigayears may well have 
transformed the nature of their AGN activity, it has recently been invoked 
(Fan et al.\ 2008) to explain another intriguing result, namely the apparent 
expansion of the effective radii of massive elliptical galaxies by a 
factor of $\simeq 3$--$4$ 
since the quasar era, as found in several independent studies (e.g., 
Ferguson et al.\ 2004; Trujillo et al.\ 2007; Zirm et al.\  2007; 
van Dokkum et al.\ 2008; Cimatti et al.\ 2008; Damjanov et al.\ 2009). 
These `superdense' ellipticals are, however, found to be exceedingly 
rare in the local universe (e.g., Trujillo et al.\ 2009).  The largest sample 
of distant galaxies used in these studies involved over 800 galaxies with 
$0.2 < z < 2$ and masses between $\sim 5 \times 10^{10}$ and $\sim 5 
\times 10^{11} M_{\odot}$ (Trujillo et al.\ 2007). These ellipticals were 
strongly argued to be much smaller at $z \ge 1.5$, with the more concentrated 
(spheroidal) galaxies having sizes roughly four times smaller 
(Trujillo et al.\ 2007) than those 
found locally in a large sample drawn from the SDSS catalogue (Shen et al. 2003). 
The less concentrated (disk dominated) galaxies also showed 
significant, albeit much more modest, size evolution for similarly high 
masses.

However a very recent alternative analysis, using a tiny but very well 
observed set of HzRGs, claims that there need be no significant 
expansion of massive ellipticals and that the multiple earlier claims 
of this important effect are in works that did not account 
for the fact that single Sersic profiles are poor matches 
to the massive ellipticals at 
low $z$ (Hopkins et al.\ 2009). These 
authors assert, firstly, that they do not find the cores of nearby 
ellipticals to be substantially different in stellar density and size, 
when compared to their high $z$ counterparts. Secondly, the absence 
of extended envelopes in the 
existing optical images of high $z$ ellipticals
(similar to those seen in low $z$ ellipticals) 
could either be a real 
effect or merely an artifact of their low surface brightness (Hopkins 
et al.\ 2009). The former possibility would be consistent with the recent 
work showing that minor mergers would expand the outer portion of the
elliptical galaxy (e.g., Naab et al.\ 2009). On the other 
hand, the latter alternative
would imply that no substantive change has occurred in the stellar 
distribution of the massive ellipticals over the past $\sim$10 Gyr 
that have elapsed 
since the quasar era. If true, this would further accentuate the difficulty
in explaining the enormous cosmic evolution observed both in the cold gas 
content of massive ellipticals and, even more, in their AGN activity (see 
above). 

If indeed the evolutionary history of RGs is associated predominantly
with a transition from a cold disc accretion to a hot accretion mode, as
mentioned above, a primary requirement would be getting rid of most of
the cold gas content of the massive elliptical hosts of the RGs after
the quasar era. The changed accretion mode would then also result in
spinning down of the central engine. Therefore, a more detailed modelling
of both these processes, namely the cold gas expulsion and BH spin-down, is
of considerable importance and the present paper will address these
two issues.  An  improved understanding of the molecular gas expulsion
mechanism has  now  acquired added significance in view of the recent
controversy over the claimed cosmological expansion of the massive 
ellipticals.

To further explore these questions, in Sect.\ 2.1 we note how 
the possible
expansion of the stellar distributions of massive elliptical
galaxies could occur via mergers and in Sect.\ 2.2 we elaborate
upon the cold gas expulsion scenario which can  lead to the very large
expansion claimed to have occurred throughout the bulk of the galaxies.
Different modes of accretion for RGs at different cosmic epochs are
considered in Sect.\ 3.
The closely related key issue of the spin evolution of the nuclear 
SMBH in massive 
ellipticals  is  then  discussed  in  Sect.\ 4. Given the current 
debate over the reality of the cosmic expansion of massive ellipticals, 
we shall treat it for the present as an unsettled issue while discussing the 
cosmic evolution of the AGN population. However, this uncertainty should 
not affect our main conclusions, which are given 
in Sect.\ 5.

\section{Galaxy expansion mechanisms}

\subsection{Galaxy expansion from galaxy mergers}

Taking the cosmic expansion of massive ellipticals to be a real effect,
Trujillo et al.\ (2007) have invoked major `dry' mergers (those in which
substantial cold gas is absent). These are expected to dominate at   
moderate redshifts over the `wet' mergers (those with 
substantial amounts of cold gas present) that could be more common at 
$z > 2-3$ (e.g., van Dokkum 2005). 
Simulations of major dry mergers yield few new stars but they 
do puff up  the effective radius of the galaxy, roughly as 
$r_e \propto M_{\star}^{0.65-1.3}$, with $M_\star$ the stellar mass 
and with the exponent in this relationship 
declining as the pericentre distance between the conjoining
galaxies increases (Boylan-Kolchin, Ma \& 
Quataert 2006). Thus, two or three nearly equal mass mergers over the past 
several billion years could account for the claimed observed 
   increase  in the galaxy size   and might also explain the 
prevailing age-uniformity 
found in local massive  ellipticals  (Trujillo et al.\ 2007).

But a recent analysis by Bezanson et al.\ (2009) strongly
indicates that major mergers face greater difficulty in producing the
 reported  large expansions and would also probably violate 
independent constraints
on mass growth provided by the evolution of the galaxy mass function.
Brighter cluster galaxies (BCGs) have been reported to evolve in size with redshift even more quickly than most of the early type population; this is most easily understood if BCGs grow from many smaller dry mergers (Bernardi 2009).
Yet another study also indicates that dry mergers have not been important
for the evolution of most ellipticals out to $z \sim 0.7$ (Scarlata et
al.\ 2007). These different claims may be consistent if results from a detailed study of Virgo cluster ellipticals (Kormendy et al.\ 2009) are generally
applicable. The brighter ellipticals in Virgo all have cuspy cores while the fainter ones do not and these coreless ellipticals have
extra light (above a Sersic profile) in their innermost portions.
That light could arise from star formation following wet mergers, while the more massive cuspy cored galaxies would have grown through
dry mergers (Kormendy et al.\ 2009). Both minor mergers and dynamical expansion due to mass loss (Sect.\ 2.2)
can yield  the  large increases in galaxy size and also accommodate 
the mass function evolution. They both could also lead to significantly 
more expansion seen in the bulk of the stellar distribution as opposed to 
just the core (within the central 1 kpc). The  ``fine tuning'' 
in the amount of 
mass  loss needed for  that scenario to be effective  may lead to a 
preference for the minor  mergers as the cause  for  most of the observed 
expansion (Bezanson et al.\ 2009).  

A detailed simulation of minor mergers 
onto a massive elliptical was recently conducted by Naab et al.\  (2009). 
They find that the dramatic increase in galaxy size, 
argued to be present by many papers over the past few  years, 
can indeed can be explained in this way. Very interestingly, this simulation 
also shows that the inner portion of the stellar distribution remains 
dominated by the stars in the original large elliptical, while the more 
extended outer reaches are mostly filled with the remnants of the smaller 
galaxies it absorbs (Naab et al.\ 2009). Another  recent study of the 
evolution of early-type galaxies indicates that variations in the times 
at which star formation halted for different galaxies  can explain 
about half of the observed expansion and
dry mergers can explain the remainder of it (van der Wel et al.\ 2009). 
Note however, that a recent very large sample of 150,000 galaxies matched 
between SDSS and the FIRST radio catalog (Becker et al.\ 1995), extending 
to look-back times of $\simeq 2$ Gyr, does not show evidence for mergers 
or other external environmental factors playing a significant role in 
triggering nuclear activity in either spiral or elliptical galaxies 
(Reviglio \& Helfand 2009). 

The merger events of the central black holes are likely  to  bring down the net spin of the new 
hole due to  the random addition of the two spins and the spin can be further reduced 
due to radiation of 
the angular momentum by  gravitational radiation (Hughes \& Blandford 2003).

\subsection{Cold gas expulsion induced galaxy expansion}

An alternative explanation for the apparently observed 
expansion of the effective radii of massive ellipticals by a factor of 
$\sim$ 3--4 since the quasar 
era invokes rapid expulsion of cold gas due to AGN feedback in the form 
of winds and jets (Fan et al.\ 2008; Sect.\ 1).  
The presence of large amounts of molecular gas in massive galaxies at 
such high redshifts is indicated by the CO detections (e.g., Papadopoulos 
et al.\ 2000; De Brueck et al.\ 2005; Klamer et al.\ 2005; Lutz et al.\ 
2008), corroborated by the result that the median redshift of sub-millimetre 
selected galaxies also coincides with the quasar era $z =$ 2--3 (e.g., 
Chapman et al.\ 2005).
Such massive galaxies are typical hosts for bright QSOs (e.g., Falomo et al.\ 2008),  and observations
of sub-mm QSOs indicate that very substantial amounts of cold gas are
present in the central regions of their hosts (e.g., Omont et al.\ 2003, Wang et al.\ 2007; Lutz et al.\ 2008).
Such sub-mm QSOs at redshifts of 2--3 are presumed to be at the transition
between very rapid star formation and the beginning of powerful outflows
launched by the luminous central engine which, however,  may still 
be shrouded by dust present in the cold gas (e.g., Hopkins et al.\ 2005).  
The amount of cold
gas found in these galaxies at such stages is comparable to the stellar 
mass, of at least the inner several kpc (Coppin et al.\ 2008; Tacconi et al.\ 2008). 

Direct spectroscopic evidence that is consistent with a very powerful outflow 
in a high $z$ RG (HzRG), with an outflow velocity of $\sim 1000$km s$^{-1}$ and
mass loss rate of $\sim 3,000 M_{\odot}$ yr$^{-1}$
has recently been reported (Prochaska \& Hennawi 2009). Likewise, 
good evidence has accumulated
for massive outflows of ionized gas from HzRGs ($z = 2 - 3)$, possibly due 
to jet feedback transferring around $10\%$ of its power to the outflowing
gas (e.g., Nesvadba et al. 2008; also, Villar-Martin et al. 
2007). Also, Tremonti, Moustakas \& Diamond-Stanec (2007) found strong outflows
exceeding 700 km s$^{-1}$ and large mass loss rates in 10 out of 14 galaxies 
with fading starbursts and AGN, albeit with $z \sim 0.6$. 
Under suitable conditions, even several times weaker 
outflows would be able to expel the cold gas in roughly one Salpeter time 
($\simeq 4 \times 10^7$ yr) from the visible portion of the galaxy (Fan et 
al.\ 2008), though the potential of the dark matter halo may well prevent 
it from being completely lost from the galaxy.
The feedback of only a small fraction of the QSO's total
peak power in winds and jets is sufficient  to yield such strong 
outflows (e.g., Granato et al.\ 2004; 
Hopkins et al.\ 2005).
Recall that powerful outflows, if produced by nearly-Eddington or even 
super-Eddington accretion onto black holes have been argued to be 
fundamentally responsible for the coordinated growth of galactic bulges 
and the SMBH mass and for  heating of the gas in clusters (e.g., Silk \& Rees
1998; King 2003; 2009).

When the mass loss is rapid compared to the dynamical time for the
stellar system then it probably causes galaxy expansion over the course of
several dozen dynamical times, an effect that has been long studied
for stellar clusters (e.g., Hills 1980). Fan et al.\ (2008) give an
approximate
calculation of the  galaxy expansion  for two limiting cases.
When the gas ejection timescale, $\tau_{\rm ej} \ll \tau_{\rm dyn}$, the
dynamical timescale, then the initial and final energies are respectively
given by $E = -\beta GM^2/2R$ and $E^{\prime} =
[-\beta GM^{\prime 2}/R + \beta M^{\prime}(GM/2R)]$, where $\beta$ depends on the
stellar density profile whose shape is assumed to be 
the same in the initial and final configurations, because neither the dispersion nor
the radius changes during the rapid  gas  expulsion.

The ratio of the new to original energies is $E^{\prime}/E =
(M^{\prime}/M)^2[2-M/M^{\prime}]$ (Biermann \& Shapiro 1979).
Therefore, if $M/M^{\prime} > 2$, the system is completely unbound but
that is not likely to ever occur for massive galaxies.  When
$M/M^{\prime} < 2$ then the system will relax to a new equilibrium
system where
\begin{equation}
\alpha(M,M^{\prime}) \equiv R^{\prime}/R = (2 - M/M^{\prime})^{-1},
\end{equation}
if the new equilibrium system is homologous to the original one.
Numerical simulations show that when the system is not disrupted the
new configuration does indeed expand by roughly this  factor  and that it
takes about 30--40 initial dynamical times for a new equilibrium to be
established (e.g., Goodwin \& Bastian 2006). In the other limit, where
$\tau_{\rm ej} \gg \tau_{\rm dyn}$, the expansion is through the adiabatic
orbits of stars,
$\displaystyle
\dot{E} = -\left (\bar{\Phi} \dot{M} +M \Delta \bar{\Phi}/\tau_{\rm dyn} \right)/2=0$, which implies 
$\displaystyle \Delta \bar{\Phi}/\bar{\Phi}= -\tau_{\rm dyn} /\tau_{\rm ej}\approx 0$,
where $\bar{\Phi}$ is the mass averaged potential and hence the expansion factor is proportional to the mass loss rate (e.g.,
Hills 1980) so, $R^{\prime}/R = M/M^{\prime}$.

Taking a reasonable value of  $\tau_{\rm dyn} \approx 5 \times 10^7$yr
in the core of a massive elliptical the timescale for relaxation should 
be $\sim$ 2 Gyr so that by $z \sim 0.8$ the compact core galaxies seen 
at $z \sim 2$ would have settled onto the fundamental plane (Fan et al.\ 
2008).
A reasonable model would adopt a Sersic profile for the projected stellar 
density, specialized to $n=4$, corresponding to a $r^{1/4}$ law (but cf.\
Hopkins et al.\ 2009),  and a 
Navarro et al.\ (1997) profile for the total mass.  With these
prescriptions and, for the sake of specificity, adopting the evolution
model of Granato et al.\ (2004),  the effective radius, $r_e$, can be
found in terms of
the virial radius, $r_{\rm vir}$, the ratio of the stellar 
dispersion to the dark matter dispersion, $f_{\sigma}$, and the masses of 
the stars $M_\star$, cold gas, $M_{\rm cold}$, and the
galaxy halo, $M_H$, (Fan et al.\ 2008),
\begin{equation}
r_e \approx \frac{0.34}{f_{\sigma}^2}~ \frac{M_{\star}+M_{\rm cold}}
{M_H}~ r_{\rm vir}.
\end{equation}
They further argue that reasonable star formation and
quasar activity inputs will produce a ratio $M_{\rm cold}/M_{\star}
\sim 2/3$ for massive galaxies  nested  in  halos with $M_H > 10^{12}M_{\odot}$.
If all of this gas is expelled in the fast ejection limit, this would
lead to an increase in galaxy size by a factor of $\sim 3$ from Eqn. \ (1),
since $M = M_{\star} + M_{\rm cold}$.
For less massive galaxies, with $M_H < 10^{12}M_{\odot}$,
the current stellar mass is $< 2 \times 10^{10}M_{\odot}$; then the
nuclear activity is probably too weak to eject the gas until  a much
later epoch, by which time $M_{\rm cold} \ll M_{\star}$ and then
$M_{\rm ej}/M_{\star}$ where $M_{\rm ej}$ is the mass of the ejected gas, 
becomes small, too. 

Now we improve upon the estimate of Fan et al.\ (2008) for $\alpha(M,M')$.  
To find the factor $M/ M'$ we note that initially 
the mass would include
the mass in the cold gas $M_{\rm cold}$; after condensation and 
star formation, part of that mass gets added to $M_\ast$. Then the
ratio of the initial to the final mass, after expulsion of the gas 
at time $t_x$, would be given by
\begin{equation}
{M \over M'}= {M_{\rm cold}(t_x) + M_{\star}(t_x) \over  M_{\star}(t_x)}.
\end{equation}
The time $t_x$  begins at the epoch when virialization
of the halo is essentially complete. 
To make further progress, we invoke the semi-analytic model of Mao et al.\ 
(2007) using the framework of Granato et al.\ (2004).  We have calculated 
this mass ratio by integrating the Eqns A1--A6 of Mao et al.\ (2007), while 
making the assumption that the expulsion event (which could be either AGN 
or supernovae driven, both of which are accommodated in the formulation) 
results in  the entire  remaining cold gas exiting from the galaxy.
With $\Delta t_{\rm burst}$ the length of star formation, the most useful 
independent variable is $\tau \equiv t_x/\Delta t_{\rm burst}$. 
Then the mass ratio is found to be
\begin{equation}
\label{tstar}
{M \over M'} =  1 +{1 -\exp{[-(s \gamma-1) \tau \delta]} \over (s-1/\gamma) 
\exp{(\tau \delta)} -s 
-\exp {\left[(s\gamma -1) \tau \delta \right]}/\gamma},
\end{equation}
where
\begin{equation}
\delta(M_{12}) \equiv {\Delta t_{\rm burst} \over t_{\rm cond}}= {5 \over 8} 
{\cal F}(M_{12}) M_{12}^{-0.2},
\end{equation}
with $M_{12}$ the total galactic mass in units of $10^{12}M_{\odot}$,
$t_{\rm cond}$ the  gas  condensation and star formation time, and
${\cal F}(x) = 1$ for $x \geq 1$ and ${\cal F}(x) = x^{-1}$ for $x \leq 1$.
The other quantities in the expression for the mass ratio are
$\gamma = 0.7 +0.6 M_{12}^{-2/3} [4/( 1+z)]$ and
$s  \equiv   t_{\rm cond}/ t_{\star} \approx 5$ for isothermal spheres,
with $t_{\star}$  defined  by the rate of star formation, $\displaystyle
 \dot{M}_\star \approx {M_{\rm cold} \over t_\star}$.
So $M/M'$ is primarily a function of redshift $z$, the total mass of the 
galaxy $M_{12}$, and $\tau$, which is the time of the  gas expulsion 
expressed in units of the starburst duration, which is taken to begin at the
time the halo became virialised (and gas condensation and star 
formation begin).  Plausible values have been adopted  for the other star 
formation and condensation related parameters (see Mao et al.\ 2007).  

\begin{figure}
\includegraphics{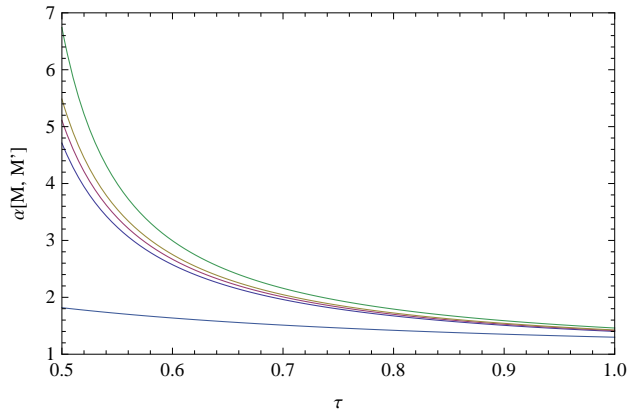}
\caption{Values of $\alpha$, the expansion factor, as functions of the
time of the  cold  gas expulsion (measured from the start of star formation in units of the  starburst duration) 
for the case of rapid expulsion 
($\tau_{\rm ej} \ll \tau_{\rm dyn}$). The four upper curves  are  for   
$z=5,3, 2.5, 2$, starting from the top. 
The lowest curve is  for the case of  adiabatic slow mass loss 
($\tau_{\rm ej} \gg \tau_{\rm dyn}$, for the case of $z=3$) where
$\tau_{\rm ej}$ is the gas exit time scale. For all curves  a halo mass of 
$M_{12} =1$  is assumed.}
\label{alpha}
\end{figure}

We would like to point out that Eqn (\ref{tstar}), without adding 1, 
is an improvement over Eqn.\ (5) of Fan et al.\ (2008) which is 
an approximation.
The deviation is significant  for low $\tau$, where additional factors
in the numerator and denominator of the expression become important. 
As a result, the
expansion factor seems to be have been underestimated in the model of Mao et al.\ (2007).
 In Fig.\ 1, the  expansion factors for the case of rapid mass 
ejection, $\alpha(\tau)$ are shown for a  range of the halo 
virialization redshift, $z=5,3, 2.5, 2$; one case of 
slow (adiabatic) mass loss is also shown in  Fig.\ \ref{alpha}. 
It is seen that for a typical value of $\tau=0.5-0.6$, and an initial 
 gas-to-stellar mass ratio of 2/3, an expansion factor of about 
3--4 is expected for the rapid  expulsion  case and  less than half 
as much for the adiabatic case. 

We see that while $\alpha > 4$ is possible it requires rather high 
$M_{\rm cold}$ to $M_{\star}$ ratios or very small values of $\tau$. 
Values of $\tau < 0.4$ imply ejection of 
much of the gas before  most  of the stars form; however, those 
circumstances  are  unlikely to yield a massive elliptical galaxy.
So while this process certainly can puff up the galaxies substantially,
it requires a large, yet not too large (otherwise the galaxy 
is disrupted), quantity of cold gas to be expelled very rapidly,  
underscoring the ``fine tuning'' problem also noted by Bezanson et al.\ (2009).

Should the expansion factor be indeed modest with $\alpha \leq 1.5$ 
(Hopkins et al. 2009) there are two  alternative possibilities for the
cold gas depletion.  First, as seen from Fig.\ \ref{alpha}, if the explosive
event is delayed by about 0.1 in $\tau$, which is 
in units of $\displaystyle  
\Delta t_{ \rm burst} =6 \times 10^8 \left[ {4 \over 1+z} \right]^{1.5} {\cal F}(M_{12})$ 
years, while the starburst has been in progress, this would lead to a lower 
$\alpha$.  There could be a variety of reasons for this: either the QSO 
switches on late or  massive stars are formed slightly later depending on 
the details of the
clumpiness of the gas distribution (we have taken it to be isothermal here). 
A second possibility envisaged is a steady adiabatic mass
loss which results in the lower curve of Fig. \ref{alpha}. 
Either of these two
situations would be consistent with the negative results of Hopkins et al.\ (2009).

\section{Cosmic evolution of the accretion mode}

While the mechanical feedback from the AGN might be responsible for the
expulsion of the bulk of cold/warm gaseous phase of   massive ellipticals during 
the quasar era (Sect.\ 2), the existing hot gas is likely to be largely
retained and even replenished by continuing stellar mass loss. Thus, 
independent of the role of  cold gas expulsion in causing 
any expansion of the galaxy, a Bondi type accretion of this hot  ISM  
phase  
onto the central SMBH could become the favored mechanism for powering 
the low-redshift/luminosity RGs (mostly of edge-darkened or  
FR I RGs), particularly because any 
emission lines seen in their optical spectra are usually of  the 
low-excitation type (e.g., Hardcastle et al. 2007; 
Balmaverde, Baldi \& Capetti 2008; also, Baum et al. 1995). 

In contrast, powerful 
RGs (FR IIs) and RQQs found at high redshifts 
are believed to be powered by a sustained thin disc accretion of cold gas, 
which is both capable of spinning up the SMBH (e.g., Rees \& Volonteri 2007; 
Berti \& Volonteri 2008) and also producing intense disc emission, in 
addition to  ejecting  powerful relativistic jets whenever conditions 
near the central engine become propitious for jet collimation (e.g., Sikora, 
Stawarz \& Lasota 2007; Livio, Pringle \& King 2003; Meier 2001). 
In contrast, the hot accretion (dominant at low $z$) should funnel only 
little angular momentum into the central BH; for such RGs the ADAF 
accretion mode (e.g., Narayan \& Yi 1995) can explain 
their observed faint disc emission with low-excitation optical spectra. 
Evolutionary scenarios along these lines are in vogue (e.g., Sommerville 
et al. 2008 and references therein), particularly since such central engines can explain the 
radiatively inefficient
``low-excitation RGs"  which characterize the bulk of the FR I 
population and some less powerful FR II sources as well (e.g., Hardcastle et al. 2007; 
Best et al. 2006; also Allen 2006). Only rare instances involving `wet' mergers at low $z$ would push 
massive ellipticals   into the high-excitation AGN mode attributable to cold gas
accretion. Evidence  now exists for the hot gas accretion to be 
the dominant process for FR I RGs across three 
decades in radio power (e.g., Balmaverde, Baldi \& Capetti 2008); that 
paper reports a nearly linear scaling between the accretion 
power and the jet power so that $\simeq$ 1\% of the rest mass energy goes 
into jet power.

Thus, it appears that the massive elliptical galaxies that had hosted 
powerful AGN
(both radio-loud and radio-quiet) during the quasar era will, in most
cases, be able to produce only their low-luminosity  versions  at small 
redshifts. We argue that observational support for massive ellipticals 
undergoing transformation from hot to cold accretion mode emerges 
from the fact that the co-moving density of powerful RGs, mostly the 
edge-brightened FR II's, peaks at 
$\phi \simeq 10^{-6.2}$ Mpc$^{-3}$ for $P_{151 \rm MHz} \simeq 10^{26.5}$ 
W Hz$^{-1}$ sr$^{-1}$ between $2<z<3$, while the weaker radio sources 
found at $z  \simeq 0$ (essentially all FR I's) have a comparable density of 
$\phi \simeq 10^{-5.8}$ 
Mpc$^{-3}$ for $P_{151 \rm MHz} \simeq 10^{24}$ W Hz$^{-1}$ sr$^{-1}$ 
(Willott et al.\ 2001; converted to $\Lambda$CDM cosmology by Grimes 
et al.\ 2004). Note that this radio luminosity is typical of FR I RGs,
being nearly an order of magnitude below the value for the most powerful 
FR I sources (Best et al. 2005). 
So, the above scheme for accretion mode change would posit that a 
substantial fraction of the  ellipticals hosting the FR I RGs at low $z$  
are probably 
descendants of the galaxies that had hosted FR II RGs at $z > 2$, as also 
hinted by Best et al. (2005).
This scenario also allows us to understand why the decline in the 
abundance with cosmic time
is so much less steep for low power radio RGs than  for  powerful RGs.  

It is further expected that hot accretion, and thus  most FR I RGs, would 
preferentially be associated with more massive galaxies at low $z$.
This is because  the Bondi accretion  rate scales approximately as
$M_{\rm BH}^{1.5}$, (e.g., Best et al.\ 2007). 
Clearly, hot accretion
would be favoured by deeper potential wells and this fits with the well
known  result  that FR I RGs are usually associated with the larger
cluster galaxies (e.g., Longair \& Seldner 1979; Best 2004; Hardcastle 
et al.\ 2007) which must also be also richer in hot ISM (e.g., Mathews 
\& Brighenti 2003). 
In the next section we discuss some important ramifications of the jet production 
that arise from the switchover to the hot accretion mode at later cosmic epoch.

\section{Jet production}

We adopt the flexible paradigm that the jets are either powered  totally by 
black hole spin, essentially by the Blandford-Znajek (1977; BZ) process
or also partially from a magnetized accretion disk, as in the 
hybrid model of Meier (1999; 2001). Note that even though jets may
also form in RQQs, they are liable to be quenched at a nascent stage 
(e.g., Gopal-Krishna, Mangalam \& Wiita 2008), 
and hence we would consider only the accretion disk output in the 
case of RQQs. The jet powers in these two cases can be written as 
(McDonald \& Thorne 1982; Meier 1999)
\begin{equation}
{\cal L}_{{\rm jet}}= 
\left \{ \begin{array}{lll} {\cal L}_{{\rm BZ}}=&{1 \over 32} \omega_F^2 
B_\perp^2 R_H^2 j^2 c &\mbox{BZ process}\\ && \\ 
{\cal L}_{{\rm H}}=&{1 \over 32 c} (B_\phi R_{ms}^2 \Omega)^2 
&\mbox{Hybrid process}\end{array}
\right.
\end{equation}
where in the first expression, $\displaystyle \omega_F \equiv \Omega_F (\Omega_H-\Omega_F)/ \Omega_H^2$ 
depends on the angular velocity of the field, $\Omega_F$, 
relative to that of the hole, $\Omega_H$, $B_\perp$ is the component of the magnetic field perpendicular 
to the hole, $R_H= r(j) m \equiv \left (1+(1-j^2)^{1/2} \right) G M_\bullet/c^2$ is the radius of the event horizon of the BH,  where $ m \equiv G M_{\bullet}/ c^2$, and  $j= a/m$
is the dimensionless angular momentum of the BH; for the second
expression, $B_\phi$ is the azimuthal component of the
magnetic field in the inner portion of the disk, $R_{ms}$ is the radius of the marginally stable orbit,
the height of the disk is taken to be $H \simeq R_{ms}$,
 and $\Omega$ is the angular velocity of the disk at 
the marginally stable orbit.

The BZ model is technically derived in the limit of zero accretion but 
the hybrid models, on the other hand, do require at least 
a modest level of accretion, most likely in the ADAF mode (e.g., 
Narayan \& Yi 1995).  As mentioned above, an ADAF type of accretion disk close to the SMBH, 
fueled by essentially Bondi accretion at larger scales, is now the
common premise for the weaker ``radio mode'' for AGN, where jets, but 
little optical emission, are produced (e.g., Somerville et al.\ 2008).  

Although some general relativistic magnetohydrodynamical (GRMHD)
simulations indicate that only an essentially BZ type of process 
is likely to yield a 
relativistic jet while hybrid type models do not work well (McKinney 2005),
other GRMHD simulations do seem to show the flows develop essentially along 
the line of the hybrid models (e.g., Koide et al.\ 2002; Nishikawa et al.\ 2005); it is fair to say that no 
consensus has yet emerged on this point. Even the BZ process for
a massive, rapidly spinning BH does not guarantee that a powerful jet will 
emerge, since it has been argued that successful formation of relativistic 
jets requires additional collimation by MHD outflows from 
accretion disks (Sikora et al.\ 2007).  When the accretion 
rate is close enough to the Eddington rate for a ``standard'' (optically 
thick but geometrically thin) accretion disc to form (e.g., Shakura \&
Sunyaev 1973), the optical and 
UV luminosity will be high, but powerful jets will only rarely emerge.

A potentially very interesting outcome of the cold gas expulsion
for both the BZ and hybrid models would arise if flux conservation 
were to be assumed for all the gas that feeds the central engine so that 
$B \propto R^{-2}$. Then the magnetic field contribution to the jet 
power, which is $\propto B^2$ in either case, is enhanced by a factor of 
about $\alpha^4 \sim 100$.  However, it is more likely that the gas 
feeding the disk, at least initially, is practically within the sole 
gravitational purview of the SMBH and therefore would not partake of 
the overall gas expulsion. This idea is supported by the observations
indicating much less stellar density evolution in the inner $\sim 1$ kpc
of massive ellipticals than in the outer portion of their stellar populations
(Bezanson et al.\ 2009).
In such an event any effect due to magnetic 
field changes is expected to be much more modest until the disk
matter is fully accreted by the BH and its spin-down sets in.

\subsection{Spin down time scale}
\label{spindown}

The formulation of the BZ model in Macdonald \& Thorne (1982) 
leads to the following forms for the jet power, 
$\cal L$, and torque,
$\cal G$ (details are given in Appendix A), 
\begin{equation}
{\cal L} = {\cal L}_0 j^2 r^2(j) ~\mbox{where}~ {\cal L}_0 ={m^2 c\over 32} B_\perp^2 g ,
\label{L0}
\end{equation}
\begin{equation}
{\cal G}=  {\cal G}_0 j r^3(j)  ~\mbox{where}~ {\cal G}_0 ={m^3 \over 8} B_\perp^2 f,  
\label{G0}
\end{equation}
where we have dropped the subscript BZ.
 The geometric factors, $g$ and $f$, are 
the results of angle averaging over the
horizon of the magnetic flux and the spin of the field and are model dependent;
expressions for them are given in Eqns (A6) and (A7). When the
maximum power is transferred from the BH to the jet, as is typically assumed,
 these factors are of order unity. 

The angular momentum budget and the rotation energy budget 
are given by, respectively
\begin{eqnarray}
{\cal J}& =& {\cal J}_0 j ~~\mbox{where}~ {\cal J}_0 = c M_{\bullet} m~~~{\rm and}\\
{\cal E}&= & {\cal E}_0 \left (1-{r^{1/2}(j)\over \sqrt{2}}\right ) ~\mbox{where}~ {\cal E}_0 =M_{\bullet} c^2   
\label{Erot}.
\end{eqnarray}
For reference we give the numerical values of the various quantities used in cgs units:
\begin{eqnarray}
{\cal J}_0 &=& c M_{\bullet} m = 9\times 10^{64} M_8^2~({\rm g ~cm^2/s}); \label{basic0}\\
{\cal L}_0 &=&{m^2 c\over 32} B_\perp^2 g =2 \times 10^{43} g B_4^2 M_8^2 ~({\rm erg/s});\\
{\cal G}_0 &=&{m^3 \over 8} B_\perp^2 f=4 \times 10^{46} f B_4^2 M_8^3~({\rm erg}). 
\label{basic}
\end{eqnarray}

This lets us compute a spin-down time from angular momentum conservation
\begin{equation}
\tau_{j,BZ}= {{\cal J}_0 \over {\cal G}_0} \int_{j_f}^{j_i} {{\rm d} j \over r^3(j) j} = 7.0 \times 10^8 ~{\rm yrs}~ 
{\left [(\kappa(j_i,j_f)/0.1) \right ]\over  B_4^2 M_9 f }
\label{tauj}
\end{equation}
where, $j_i$ and $j_f$ are the initial and final spins, respectively, 
$\kappa(j_i,j_f)$ is the value of the integral, $B_4 = B/10^4$ Gauss
and $M_9 = M_{\bullet}/10^9 M_{\odot}$.
The spin down time
can be found by calculating the time for the rotational energy to reduce
by a factor $\epsilon$; see Fig.\ \ref{tau}. A detailed evaluation of this factor is given in Appendix \ref{spin}. The time corresponding to $\epsilon=1/e$ with  $j_i= 0.5$
is calculated to be 0.5 Gyr for $M_9=1, B_4=1$.

\begin{figure}
\includegraphics{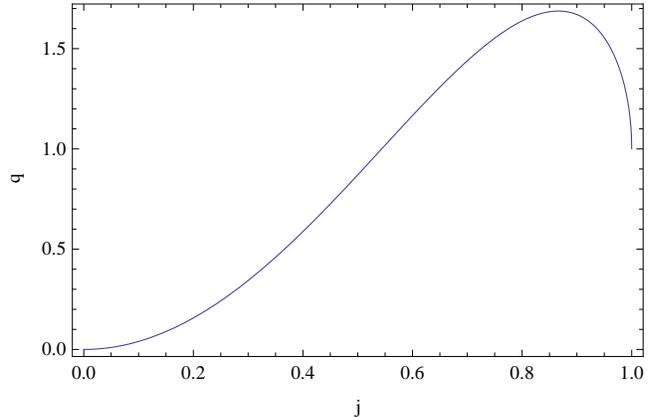}
\caption{The power of the BZ jet as a function of the BH spin parameter $j$, normalized to the
value at $j=1.0$.
The power goes through a maximum at $j=\sqrt{3}/2$ before dropping; this arises from the competition between the
horizon radius and the spin of the BH.}
\label{lum}
\end{figure}

\begin{figure}
\includegraphics{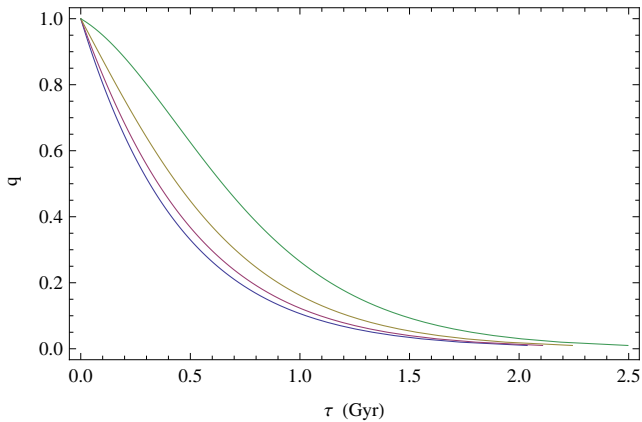}
\caption{The power reduction factor, $q$, of the BZ jet as a function of time
is shown for various initial values of the BH spin parameter $j=0.2, 0.4,0.6,0.8$ from bottom to top. The values assumed for other parameters are $B_4=1, M_9=1$.}
\label{power}
\end{figure}

Another estimate of the spin down time that is more relevant observationally is the e-folding time scale of the jet power. 
The $j_f$ at a time when a reduction by a factor $q$ is reached can be calculated from 
\begin{equation}
q j_i^2 r^2(j_i) = j_f^2 r^2(j_f).
\label{q}
\end{equation} 
The above equation leads to a quartic in $j_f$,
\begin{equation}
j_f^4 -2 q j_f +q^2=0, 
\end{equation}
whose roots are expressible analytically. 
The positive real root between $0$ and $j_i$ is found to be $j_f(q,j_i)$, 
which is fed into Eqn (\ref{kappa})
to calculate the time in terms of $\tau_j$ from Eqn (\ref{tauj}).
The ${\cal L}(j)$ goes through a maximum at $j=\sqrt{3}/2$ (Fig.\ \ref{lum}). The evolution of the
power as a function of time is given in Fig.\ \ref{power}.
 The time corresponding to $q=1/e, j_i= 0.5$
is calculated to be 0.5 Gyr for $M_9=1, B_4=1$. The spin down
time scale is likely to be reduced further when the mass of the hole increases by Bondi accretion 
(with negligible net angular momentum) as it spins down. The effective timescale is estimated to be 
$\displaystyle \tau_{spin} \approx {(\tau_a/2) \tau_j \over (\tau_j +\tau_a/2)}$, 
where the accretion time scale is $\displaystyle \tau_a \equiv {M_\bullet c^2 \over L_E} =$0.45 Gyr,   
where $L_E$ is the Eddington luminosity. As a result, $\tau_{spin} \approx 0.2$ Gyr
for the typical case $j_i=0.5, M_9=1, B_4=1$.
Note that the timescales derived above are inversely proportional to
the BH mass (Eqn \ref{tauj}).

\begin{figure}
\includegraphics{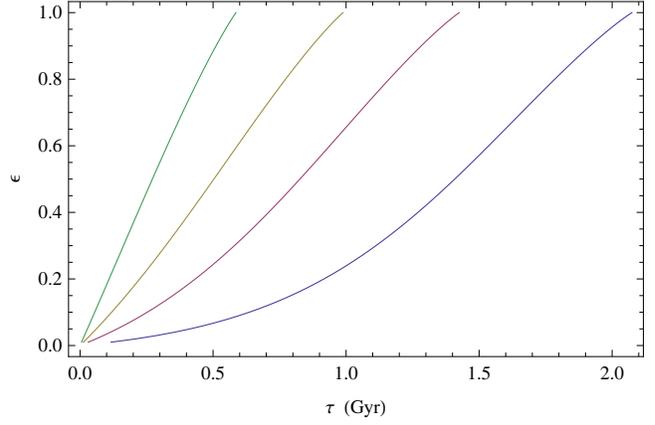}
\caption{The fraction of BH rotational energy, $\epsilon$, lost by the BZ process as a function of time
is shown for various initial values of the BH spin parameter $j=0.2,0.4,0.6,0.8$ from right to left. The values assumed for other parameters are $B_4=1, M_9=1$.}
\label{tau}
\end{figure}

The energy loss time-scale is quite similar for the hybrid model,
although the dependence on BH mass is much more gradual.  
Using the rotational energy from Eqn (\ref{Erot}) above in the 
limit of low spin, 
and Eqn (12) for the power in the MHD jet given by Meier (1999), 
we have the time scale
\begin{equation}
\tau_{L,hybrid} = 5 \times 10^8 {\rm yr} ~\alpha_{-2}^{-1/10} M_9^{-1/10} {\dot m}_{-3}^{-4/5} \zeta_{-1}^{-1},
\label{hybrid}
\end{equation}
where $\alpha$ is the disk viscosity parameter, 
${\dot m}$ is the accretion rate
in units of the rate that would produce an Eddington luminosity, and $\zeta$ is a duty cycle parameter expected to be around $10^{-1}$ (Meier 1999). All 
quantities are normalized to the power of ten that is indicated by the subscript
 and are chosen to be typical for an ADAF type of flow.  Thus we find that 
in either case, for a BH of mass about $\sim 10^9$M$_{\odot}$ the 
time-scale for spin-down is a few times $10^8$yr. 
Note that in the hybrid model the spin-down time scale is extremely
insensitive to $M_{\bullet}$ (Eqn \ref{hybrid}), while it scales as $M_{\bullet}^{-1}$
for the BZ model (Eqn \ref{tauj}) where the
angular momentum is directly extracted from the hole as opposed to the hybrid case, where it is partly extracted from the accretion disk. 
An important caveat to note is that the spin-down time scales derived above (eg. Eqn (\ref{tauj})) depend on the assumption that the magnetic field strength is constant during the spin-down. 

\section{Discussion and Conclusions}

In this paper we have focused attention on the transition period 
connecting the quasar era $(z = 2 - 3)$ to the present era, between
which the co-moving space density of powerful radio galaxies and QSOs 
has declined by almost three orders of magnitude. As described in
the foregoing sections, this ``transitional" era is probably marked  
by a changeover in the dominant mode of AGN activity from one fuelled
by accretion of cold gas onto the SMBH to another which is dominated 
by Bondi accretion of hot gas from the interstellar medium of the
galaxy (Sect. 3, Somerville et al.\ 2008 and references therein). 
Only the cold accretion mode can support a strong disc emission 
and a high spin of the central supermassive black hole (SMBH) which
can therefore eject powerful jets under suitable conditions.
An important consequence of this paradigm of accretion mode change 
is that the massive elliptical galaxies that were able to host powerful 
AGNs at high redshifts would usually be no longer able to do so at recent cosmic
epochs (Sect. 3).
 
The postulated basic operational flip of the central engines in
massive ellipticals since the quasar era underscores the need to 
take a closer look at the physical processes governing the transitional 
era. Clearly, the two dominant processes underlying this 
transformation are the elimination of vast reservoirs of 
molecular gas present in the massive ellipticals at high-$z$ and 
the consequent spinning down of their SMBH. In this paper we have 
addressed both these points quantitatively by presenting an improved 
theoretical modelling of the physical mechanisms involved.

Based on the available evidence we have argued in favour of the AGN 
wind or jet driven expulsion as being the principal mechanism by which 
massive ellipticals have lost their vast reservoirs of molecular gas 
since the quasar era. Recently, this issue of cold gas ejection has 
attracted a great deal of attention, following the claims of an observed 
$\sim 3-4$ fold increase in the sizes of massive ellipticals since the 
quasar era, along with a very recent counter-claim (Sect.\ 1).

We have obtained a detailed, improved expression for the expansion factor, 
Eqn (\ref{tstar}), which
allows us to distinguish between the expulsion histories of galaxies in terms of the
star formation parameters. The distinction is expressed in our model 
as the  expansion factor would be
lower if more stars form in the core before the gas expulsion process begins. 
Alternatively, a steady adiabatic 
mass loss leads to a modest expansion factor. In the former case, 
the core would be more luminous as more star formation occurs and less gas 
is expelled. More detailed modelling along these lines 
can explore the space of the star formation parameters, the timing of gas 
expulsion, and the cases between adiabatic and rapid mass loss, in order 
to make better connections with observations.

Irrespective of the potential role in causing the reported dramatic
size evolution of massive elliptical galaxies, the cold gas expulsion 
will only have a delayed effect on the output of the central engine 
(accretion disk plus the SMBH). This is because the cold gas associated 
with the nuclear gas cloud that is gravitationally bound to the SMBH 
is expected to be only weakly subject to the AGN induced cold gas 
expulsion. This resulting residual AGN activity from the cold accretion era would 
largely define the transitional era and be itself determined by the 
duration of the spin-down phase which the SMBH would trace once the 
nuclear molecular cloud is fully accreted and, consequently, a dramatic 
fading of the central engine has set in.

We have focused largely on the Blandford-Znajek mechanism of extracting 
rotational energy from the BH in the absence of accretion torque. 
We have derived two time scales
using energy loss and power reduction as criteria for determining the spin
down, which turns out to be about $0.5/(B_4^2 M_9)$ Gyr. If Bondi accretion proceeds,
it will add to the mass of the BH but little to its angular momentum and 
thus reduce the spin-down time scale to about $0.2/(B_4^2 M_9)$ Gyr. The evolution of 
the jet power indicates
an increase before a gradual decline if the initial spin, 
$j>\sqrt{3}/2$, as a result of the hole's increasing size. 
This naturally has implications for the evolution of the jet.  
We plan to expand our work to
hybrid models in greater detail, and to thus explore disc accretion models 
that explicitly involve angular momentum transport 
from the hole to the disc. 

The transition from cold to hot accretion dominated phase in the cosmic 
evolutionary history of the AGN population is marked by a period when 
the SMBH would continue to be fuelled (and spun up) by the accretion of the
cold gas located within the sphere of influence of the SMBH. 
This gas is likely to survive the otherwise efficient cold gas expulsion due 
to the intense AGN activity during the quasar era. But, even if such nuclear 
gas cloud has
a radius as large as 1 pc, it would sustain cold disc accretion phase for 
no more than $10^9$ years (e.g., King \& Pringle 2007). Moreover, the 
spin-down of the central engine would also occur in $\sim 10^9$ years (also,
Meier 2001) for reasonable values of mass accretion rate and the duty cycle 
parameter (Eqn \ref{hybrid}). Thereafter, i.e., during the past 6-7 Gyr, the 
occurrences of cold accretion dominated AGN would become rare, mostly 
sustained by occasional `wet' mergers of captured galaxies. Thus, the 
AGN activity over the past several gigayears would be increasingly marked 
by Bondi accretion of hot gas powering the central engines and thereby 
producing mostly FR I RGs (Best et al. 2006; Hardcastle et al. 
2007).

To summarize, we have envisaged a scheme characterized by the following
sequence of events related to the evolution of massive elliptical galaxies. At $z \simeq 3$
they contain an abundant supply of cold gas that yields both prolific 
star formation and luminous thin disk accretion. The latter usually 
leads to a fast spinning BH.  Once a good fraction of the cold gas has 
been accreted, the resulting fast spinning BHs would become capable of
of ejecting powerful jets and forming the luminous radio sources that 
were so much more abundant during the quasar era. But these powerful jets 
and disc winds can easily expel most of the cold gas reservoir, possibly
resulting in a substantial expansion of the host galaxy. A likely outcome 
of the cold gas expulsion is the eventual dramatic weakening of the 
central engine itself through a spinning down of the SMBH, usually on a timescale 
less than $\sim 1$ Gyr.

We thank the anonymous referee for suggestions that improved the presentation of our results.
G-K and P.J.W.\ thank the Indian Institute of Astrophysics for the local hospitality
provided during their visits and P.J.W.\ also thanks NCRA for hospitality. P.J.W.\ is supported in part by a subcontract from
NSF grant AST 05-07529 to the University of Washington.

\appendix
\section{Formulae for power and torque}
\label{macthorne}
We derive the Eqns (\ref{L0}--\ref{basic}) given in Sect.\ \ref{spindown}. We start
from Eqn (7.19) in Macdonald \& Thorne (1982) for the luminosity through a ring
at latitutde $\theta$
\begin{equation}
\Delta {\cal L}= {\Omega_F (\Omega_H-\Omega_F) \over 4 \pi c} \varpi^2 B_n \Delta \psi,
\end{equation}
where $\varpi(\theta)= R_H(j) \sin{\theta}$ is the ring radius, $\Omega_H(j)$ is the angular
velocity of the hole, $\Omega_F$ is the angular velocity of the field, $B_n={\bf B \cdot n}$ is the
field perpendicular to the hole surface and $\Delta \Psi (\theta)$ is the flux through a ring at $\theta$ of
arc length $R_H(j) \Delta \theta$, which is given by
\begin{equation}
\Delta \psi (\theta) = 2 \pi \varpi(\theta) R_H(j) B_n \Delta \theta.
\end{equation}
The torque over the same ring is given by
\begin{equation}
\Delta {\cal G}= {\Delta {\cal L} \over \Omega_H -\Omega_F}.
\end{equation}
Now when the above equations for the  power and torque are integrated over  $\theta$ 
we obtain
\begin{equation}
{\cal L}= {B_\perp^2 R_H(j)^4 \Omega_H(j)^2 \over 8 c} g ~~~{\rm and}
\end{equation}
\begin{equation}
{\cal G}= {B_\perp^2 R_H(j)^4 \Omega_H(j) \over 4 c} f,
\end{equation}
where the geometric factors factors $f$ and $g$  
are found to be
\begin{eqnarray}
f&=&\int_0^\pi \left ({({\bf B \cdot n})^2 \over B_\perp^2}\right )\left ( {2 \Omega_F \over \Omega_H} \right)  
\sin^3{\theta}~ {\rm d} \theta  \\  
g&=&\int_0^\pi \left ({({\bf B \cdot n})^2 \over B_\perp^2}\right ) \left ( {4 \Omega_F \over \Omega_H} \right)
\left ( 1-{\Omega_F \over \Omega_H} \right) \sin^3{\theta} ~{\rm d} \theta,    
\end{eqnarray}
where ${\bf B}$ and $\Omega_F$ could be functions of $\theta$ and 
$B_\perp^2 \equiv \int_0^\pi ({\bf B \cdot n})^2 \sin^3{\theta}~{\rm d} \theta$ is
the angle averaged root mean squared value of the normal component of the field 
on the surface of the hole.
These factors are of order unity; if the maximum power is transferred,
then $\Omega_F= \Omega_H/2$, and both of these integrals are unity. Then  Eqns (\ref{L0}, \ref{G0})
follow after substituting for $R_H(j)$ and $\Omega_H(j)$, which are respectively
the black hole's radius
and angular velocity and are given by
\begin{eqnarray}
R_H(j)&= &r(j) m =\left (1+(1-j^2)^{1/2} \right) G M_\bullet/c^2, \\
\Omega_H(j)&=& \left ({j \over 2} \right)  { c \over R_H(j)}. 
\end{eqnarray}
Eqns (9) and (10) then follow as the fundamental equations for the black hole angular momentum and
rotational energy in terms of the irreducible mass of the hole. The coefficients for these
physical quantities and the luminosity are easily found to be Eqns (\ref{basic0}--\ref{basic}). 

\section{Calculation of the spin down factor}
\label{spin}

We evaluate the energy radiated when the spin reduces from $j_i \rightarrow j_f$ to be
\begin{equation}
\Delta {\cal E} =\int {\cal{L}} {\rm d}t= {\cal L}_0 \int r^2(j) j^2 \left[ {{\rm d } j\over {\rm d }t} \right ]^{-1}
{\rm d }j
\end{equation}
As a fraction of the rotational energy budget, this is found to be
\begin{equation}
\epsilon(j_i,j_f) \equiv {\Delta {\cal E} \over {\cal E}} =  \left [ \int_{j_i}^0 
{j \over r(j)} {\rm d} j \right ]^{-1} \int_{j_i}^{j_f} {j \over r(j)} {\rm d} j
\label{eps}
\end{equation}
The integral $\kappa$ defined in Eqn (\ref{tauj}) is calculated to be
\begin{equation}
\kappa(j_i, j_f) = \left[ \left ({1 \over 16} \right) \ln{\left ( {2-w \over w} \right)}
+ \left ( {3 w^2 +3w -4 \over 24 w^3} \right) \right ]_{w_f}^{w_i}
\label{kappa}
\end{equation}
where $w_i=r(j_i)$ and $w_f=r(j_f)$. Using similar techniques the integral 
\begin{equation}
\chi(j_i, j_f) \equiv \int_{j_i}^{j_f} {j \over r(j)} {\rm d} j = \left [ \ln{(w)}-w \right ]_{w_i}^{w_f}.
\label{chi}
\end{equation}

Now from Eqn (\ref{eps}), the value of $w_f$ for $\epsilon=1/e$ and $j_i= 0.5$ is solved for and fed into
Eqn (\ref{kappa}).  Then using Eqn (\ref{tauj}) the time to spin down is found to be $\simeq 0.5$ Gyr.

\end{document}